# Creation of radiocarbon C-14 at thunderstorms

V I Lyashuk [1,2]

[1] Institute for Nuclear Research of the Russian Academy of Sciences, Moscow, Russia
[2] National Research Center "Kurchatov Institute", Moscow, Russia

**Abstract**. The synthesis of isotopes is possible under conditions of power electric discharge in the atmosphere. Knowledge of the radioactive $^{14}$C yield under flash conditions (as additional channel of $^{14}$C production relative to the main - cosmogenic one) is important for radiocarbon analysis. It is proposed the gross model for evaluation of the upper limit of the $^{14}$C yield, which creation was simulated for the altitudes up to 15 km. It is presented the results for yield of radioactive isotope $^{41}$Ar which synthesis goes along with $^{14}$C creation under thunderstorm conditions. It was obtained that the possible thunderstorm mechanisms of $^{14}$C creation cannot compete with cosmogenic production.

**1. Introduction.**
The main mechanism of radiocarbon $^{14}$C creation on the Earth is ensured by cosmogenic irradiation [1] with yield of 472 g-mole/year in the reaction of thermal neutrons with atmospheric nitrogen: $^{14}$N(n,p)$^{14}$C. The generated isotope of $^{14}$C is assimilated in the biomass (in the form of dioxide $CO_2$) and decays within it ($T_{1/2}$ =5700 y) that allows to date the age of the investigated organic materials.

Along with such cosmogenic generation it is possible the synthesis of the isotope $^{14}$C under conditions of atmospheric thunderstorm: the electrons in the avalanche of the flash discharge slow down and escape bremsstrahlung *x*-rays; the escaped *x*-radiation produces the flux of photo-neutrons in interactions with air nitrogen in $^{14}$N(γ,Xn)-reaction ($E_{threshold}$ = 10.6 MeV), where Xn – emission of X=1, 2 or more neutrons with maximal cross section at $E_γ ≈ 23$ MeV according to JENDL-3.3 nuclear reaction library [2]); the produced neutrons slow down and intensively create the radiocarbon in reaction $^{14}$N(n,p) $^{14}$C (cross section according to ENDF/B-VIII [3] is given in Fig.1; thermal cross section $σ_{np} ≈ 1.8$ b).

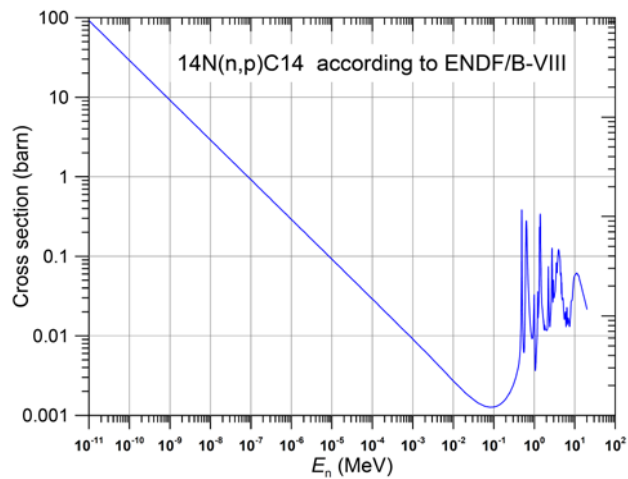

**Figure 1.** Cross section of radiocarbon C-14 production in $^{14}$N(n,p)$^{14}$C reaction according to the nuclear reaction library ENDF/B-VIII [3].

As neutron production is the threshold process, then the $^{14}$C synthesis can be realized only by means the relativistic electrons with energy higher than threshold of the reaction. The spectrum of these energetic electrons was applied as $f \sim \exp(-\varepsilon[\text{MeV}]/7.3)$ [4], where $\varepsilon$ is the energy of the runaway electrons (i.e., electrons accelerated in the electric fields; the process was investigated by Wilson [5]). The spectrum spreads up to ~60 MeV ensuring the multiplication of the avalanche under condition of atmospheric electric fields in the thunderclouds. Namely relativistic electrons ($E >1$ MeV) move in the forward part of the flash discharge producing the low energy electrons in interactions (via ionization of the media), drawing them into the avalanche propagated and accelerated in the thundercloud electric field. In opposite the electrons which energy decreases below the threshold about 100 eV are fall out from the avalanche and form the dynamical equilibrium between involved and lost electrons. In the avalanche the number of low energy electrons $N_{le}$ strongly exceed the relativistic ones $N_{re}$, the relation is $N_{le}/N_{re} \approx (1.3 \times 10^4) \times n$, where $n = \rho(H)/\rho(H=0)$ is the relation of the air density $\rho(H)$ at the altitude $H$ to the density at the sea level ($H = 0$) [4]. So, the total charge of flashes is ensured namely by low energy electrons which part in the avalanche decreases for higher altitudes.

## 2. The proposed gross model for simulation

Taking into account the dependence of relation $N_{le}/N_{re}$ from the air density the simulation was realized for the altitudes from the sea level up to the $H=15$ km (i.e., including approximately the upper charge layer at typical elevation $H=(10-14)$ km) as the most of thunderclouds are distributed at these heights [6]. We use the spherical geometry with centers (the point source of energetic electrons of isotropic $f$-spectrum) at the indicated altitudes. The spheres are divided into plane layers (of 500 m thickness) with air density corresponding them heights. In order to exclude the escape of the valuable part of neutrons (which were born in the sphere) the radii were increased up to 30 km. The scheme of geometry is given in the Fig. 2.

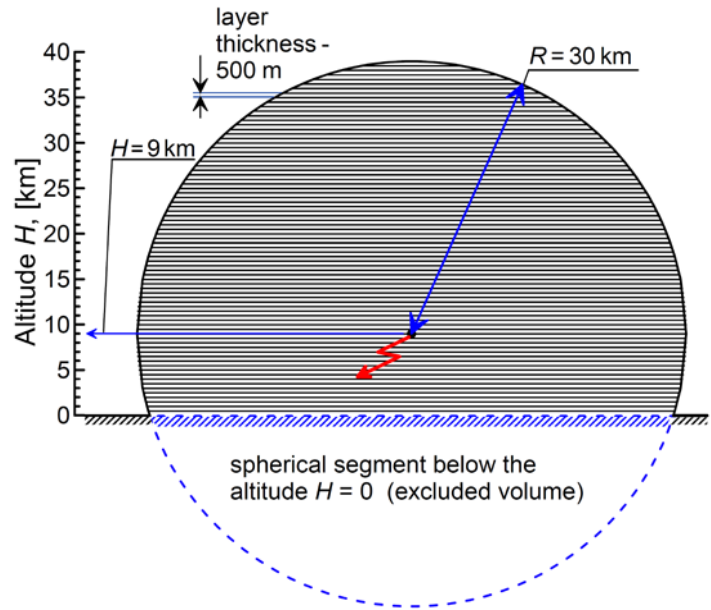

**Figure 2.** Geometry of the spherical-layers model for simulation of the particles transport and calculation of radiocarbon $^{14}$C creation in the air under conditions of thunderstorms lightning (examples for lightning [indicated as red arrow] origin at the altitude $H = 9$ km on the sea level). The spherical segment below the sea level ($H =0$) is excluded from $^{14}$C accumulation.

As a result the percent of the escaping neutrons was lower than 1%. An example of the spectrum of generating neutrons at $H$=10 km is presented in the Fig. 3 (a). The maximum of the obtained spectrum ~ 23 MeV in Fig. 3 (a) is in good agreement with maximum of neutron production in $^{14}N(\gamma,Xn)$ reaction in Fig. 3 (b). Such a spherical-plane-layers formalism allowed to specify the yield of relativistic electrons $N_{re}$ in the total $(N_{le} + N_{re})$ – flux that was necessary for correct evaluation of $^{14}C$ production.

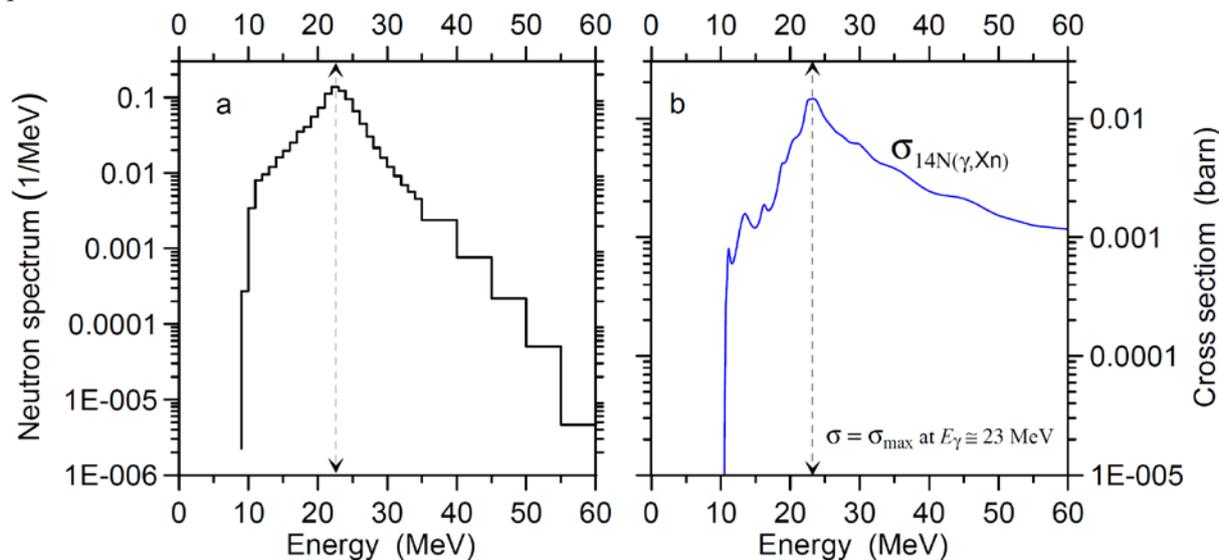

**Figure 3.** (a) Spectrum of neutrons produced in air under thunderstorm conditions. Generation of neutrons is accounted in the reactions: $^{14}N(\gamma,Xn)$, $^{16}O(\gamma,n)$ $^{15}O$ and $^{40}Ar(\gamma,n)^{39}Ar$. (b) Cross section of the main channel of neutron production $^{14}N(\gamma,Xn)$ (according to JENDL-3.3 [2]) by bremsstrahlung from electrons under thunderstorm flashes. The spectrum maximum is in good agreement with the maximum cross section of the main channel of neutron production $^{14}N(\gamma,Xn)$ (see (a)).

The modeling results (by means the code [7]) for radiocarbon $^{14}C$ yield depending on the altitude are shown in the Fig. 4.

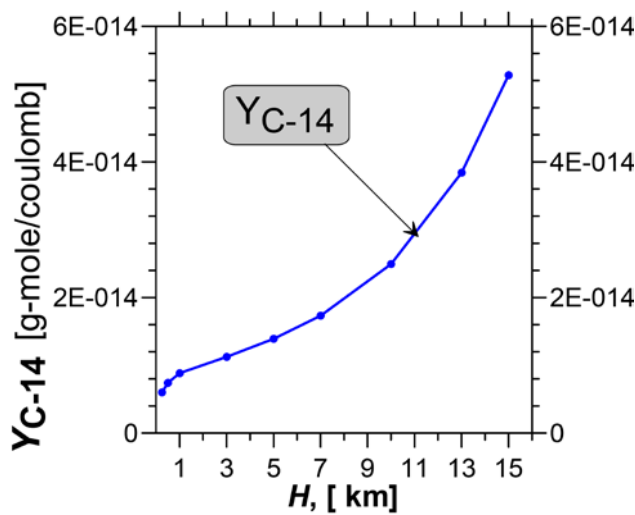

**Figure 4.** Yield of $^{14}C$ (in gram molecules) depending on the altitude $H$ (km) above the sea level. The yields correspond to one coulomb flash discharge under thunderstorm conditions.

For the equal discharges the drop of the low energy population $N_{le}$ in the avalanche at increase of the altitude ensures rise of the $^{14}$C yield. The results of isotope generation (in gramme-molecules) are normalized on the flash charge 1 coulomb. If the discharge occurs between thunderclouds in the horizontal plane (idealized case) then the normalized yield corresponds the function *Y(H)* in the Fig. 4. In common case the discharge goes between some altitudes *H1* and *H2*. The $^{14}$C yield is calculated then as the integral along the discharge path and normalized yield will be between *Y(H1)* and *Y(H2)*.

## 3. Evaluation of the upper limit for radiocarbon C-14 production. Creation of Argon-41 under thunderstorm conditions

Let us evaluate the upper limit of $^{14}$C production per year under the flash condition (knowing the number of flashes on the Earth per 1 year – $1.4 \times 10^9$ [8] and considering that the average flash charge is ~ 20 coulombs [6] and mean $H = 7$ km [6]): then production $Y_{C-14} = 1.7 \times 10^{-14} \times 20 \times 1.4 \times 10^9 \simeq 5 \times 10^{-4}$ (g-mole/year) for the relation *R1*.

Creation of neutron flux ensure simultaneously also production of $^{41}$Ar isotope by the reaction $^{40}$Ar(n,γ)$^{41}$Ar with significant cross section (see Fig.5).

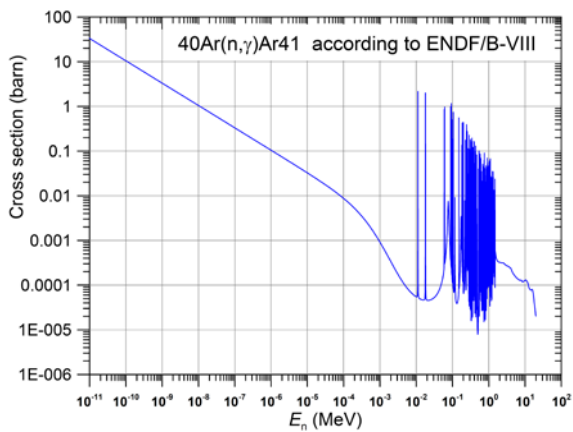

**Figure 5.** Cross section (barn) of the $^{40}$Ar(n,γ)$^{41}$Ar reaction according to ENDF/B-VIII library.

In the similar way (as $^{14}$C production) it was obtained the yield of radioactive $^{41}$Ar (produced at neutron activation of the main isotope of argon: $^{40}$Ar(n,γ)$^{41}$Ar) – see Fig. 6. The upper limit of $^{41}$Ar production per mean flash (20 coulomb) will be: $Y_{Ar-41} = 2.9 \times 10^{-17} \times 20 \simeq 5.8 \times 10^{-16}$ (g-mole).

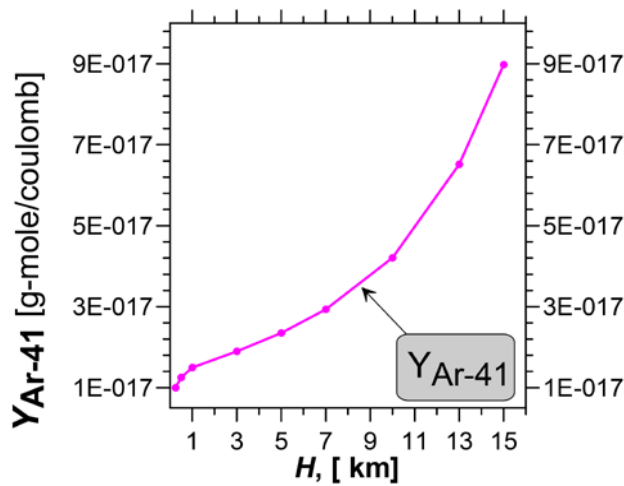

**Figure 6.** Yield of $^{41}$Ar (gram molecule) depending on the altitude *H* (km) above sea level. The yields correspond to one coulomb flash discharge under thunderstorm conditions.

The curves of yields for $^{14}$C and $^{41}$Ar are very similar as: the main yield occurs at the thermal energy; $(N_{le}/N_{re}) \propto n$. As a result, $Y_{C-14, Ar-41} \propto \sigma_{C-14, Ar-41} \times n$. Note that cross sections for $^{14}$C and $^{41}$Ar production $\sigma \propto 1/v$ (at $E < 2 \times 10^{-2}$ MeV for $^{14}$C and E $< 1 \times 10^{-4}$ MeV for $^{41}$Ar; $v$ – neutron velocity)

The obtained upper limits for yields of $^{14}$C and $^{41}$Ar were evaluated basing on the relation $N_{le}/N_{re} \approx 1.3 \times 10^4$. According to the alternative model of electron avalanche this relation is more than factor of order larger: $N_{le}/N_{re} \approx 3 \times 10^6$ [9]. In case of this relation the yields for $^{14}$C and $^{41}$Ar will be smaller in $(3 \times 10^6/1.3 \times 10^4)$ times. Taking into account two alternative theories for evaluation of the possible upper limits of $^{14}$C and $^{41}$Ar creation we have to use the first $N_{le}/N_{re}$ relation.

Under thunderstorm condition the isotope $^{41}$Ar is created in the atmosphere simultaneously with $^{14}$C. Owing to relevant $^{41}$Ar decay characteristics ($T_{1/2}$ = 109.34 m, $\beta^-$(100%)) it will be attractive to the consider this isotope as an appropriate tracer of the radiocarbon $^{14}$C generation. But the detection of such low and changing 41Ar concentration is a very complicated task. In spite of debugged technique of $^{41}$Ar monitoring (for example on the accelerators [10] and reactors) the detection of $^{41}$Ar may be possible in cases of significantly larger atmospheric discharge phenomena (may be when population of neutrons reaches ~ 1E+15 as in case of large terrestial gamma flashes [11]).

**4. Linear and mass collision stopping power and bremsstrahlung**

In the previous part we have investigated the dependence of isotope yields from the altitude, that means that yields are sensitive to the air density. But the isotope composition of the atmosphere is not vary with the height. So, the probabilities of the reaction channels does not depend on the height and we can expect that yields of isotopes will be independent on the altitude. The mechanism of this dependence of isotope yields from the altitude is the next: it is created by increase of relativistic electron part $N_{re}/(N_{re} + N_{le})$ in the avalanche at rise of the altitude.

But exists also an another mechanism in which the probabilities of (γ,n)-reaction (for $^{14}$N(n,p) $^{14}$C, $^{40}$Ar(n,γ)$^{41}$Ar production) is sensitive to the air density. These processes is connected with stopping power of electrons and bremsstrahlung. Let us consider and evaluate the significance of these processes for our task of isotope production under thunderstorm conditions.

<u>Slowing down of electrons</u>. In inelastic interactions the mean energy loss per unit of pathlength is called as stopping power. The stopping power is given as:

$$-\frac{dE}{dx} = \mathbb{N} \int W \frac{d\sigma}{dW} dW, \qquad (1)$$

where $\sigma$ - cross section of inelastic scattering with $W$-energy loss; $\mathbb{N}$ - concentration of the scattering centers in unit of volume. The stopping power is the unit, which characterizes the rate of ionization process and excitation of the matter at slow down of the charged particle. dE/dx – is called as linear stopping power. The above defined stopping power is called as collision stopping power $S_{collision}$ [12]. For the considered here electron energy [~(10-60) MeV] the main yield to the energy loss is occur in ionization and expressed by Bethe formulae [13, 14].

In order to exclude the dependence on the matter density the rate of energy loss is given as (dE/dx)/$\rho$, (MeV×cm$^2$/g); where $\rho$ - density of the matter. (dE/dx)/$\rho$ – is called (in terminology of ICRU Report [12]) as mass collision stopping power and denoted as $S_{collision}/\rho$:

$$\frac{S_{collision}(T)}{\rho} = \frac{2\pi r_e^2 mc^2}{\beta^2} N_A \frac{Z}{A} \left\{ \ln(T/I)^2 + \ln(1+\tau/2) + F(\tau) - \delta \right\}, \qquad (2)$$

$$F(\tau) = (1-\beta^2)\left[1 + \tau^2/8 - (2\tau+1)\ln 2\right], \qquad (3)$$

where $T$ – kinetic density, $r_e$ – classical electron radius ($2.8179 \times 10^{-13}$ cm), $mc^2$ – the rest energy of the electron (0.511 MeV), $\beta$ – electron velosity (in units of light velosity), $N_A$ – Avogadro number, $A$ – atomic weight, $Z$ – the atomic number, $\tau$ – kinetic energy (in $mc^2$ units), $I$ – mean value of excitation energy (obtained as a rool from experimental data), $\delta$ - density effect correction (the parameters reduces the nuclear Coulomb forces at polarization of the media at moving of charged particles, density effect correction takes into account the dielectric properties of the media).

<u>Bremsstrahlung</u>. The Coulomb field of the nuclei decelerates the velocity of the charged particles: as a result the charged particle irradiates gamma quants [13,14].

The energy loss due to bremsstrahlung per unit of the path (normalized on the matter density $\rho$) is called as radiative stopping power:

$$-\frac{1}{\rho}\left[\frac{dE}{dx}\right]_{radiative} = \frac{N_A}{A}\alpha r_e^2 Z^2 \left(T_1 + mc^2\right)\Phi_{raduative} \quad , \tag{4}$$

where $T_1$ – incident electron kinetic energy (in $mc^2$ units), $\alpha$ - the constant of the fine structure.

$$\Phi_{radiative} = \int_0^{T_1} k\frac{d\sigma}{dk}dk \bigg/ \left[\alpha r_e^2 Z^2 \left(T_1 + mc^2\right)\right] \quad , \tag{5}$$

$k$ – photon energy, ($T_2$ - electron energy after irradiating of the photon, $T_2 = T_1 - k$).

According terminology of ICRU Report 37 [12] the (normalized on density) radiative stopping power is called as the mass radiative stopping power and denoted as $S_{radiative}/\rho$.

The relation of ionization loss to bremsstrahlung radiation is:

$$\frac{S_{collision}(T,\delta)}{\rho} \bigg/ \frac{S_{radiative}(T)}{\rho} \quad , \tag{6}$$

where the density value $\rho$ in ionization and bremsstrahlung loss is cancelled out. But in spite of cancel out of the density $\rho$, the weak dependence on matter density exists in $S_{collision}(T,\delta)$ due to correction on density effect $\delta$.

In gases the values of density effect $\delta$ is significant [about (several units)×0.1 at the pressure 1 atmosphere] and increases with energy and pressure rise. The density effect correction $\delta$ was calculated according to the work [15] for the relativistic electron energy in avalanche in air at different altitudes – up to 10 km (see Fig. 7).

Now let us evaluate the significance of density effect correction $\delta$ for yield of bremsstrahlung (we remember an importance of bremsstrahlung for ($\gamma$,n)-process and for the next step: isotope production under the neutron flux) in case of altitude elevation. Note again that normalized collision stopping power depends on the altitude, but the normalized bremsstrahlung does not depend (see Eq.(6)).

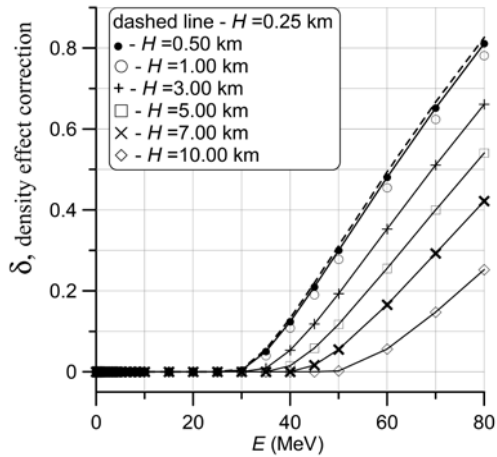

**Figure 7.** Density effect correction $\delta$ for electrons in the air for different altitudes $H$ above sea level.

For relativistic electrons in the avalanche the calculated results for the two main energy loss processes (stopping power and bremsstrahlung) are given in the Fig. 8. The results demonstrate that at the upper energy of electrons in the avalanche (~60 MeV) the stopping power are practically equal. Even at the energy 100 MeV the difference in stopping power at $H$=0 and $H$=10 km is small (the corresponding values are 2.42 and 2.47 MeV×cm$^2$/g). So the change in bremsstrahlung yield in the total energy loss will be very small too.

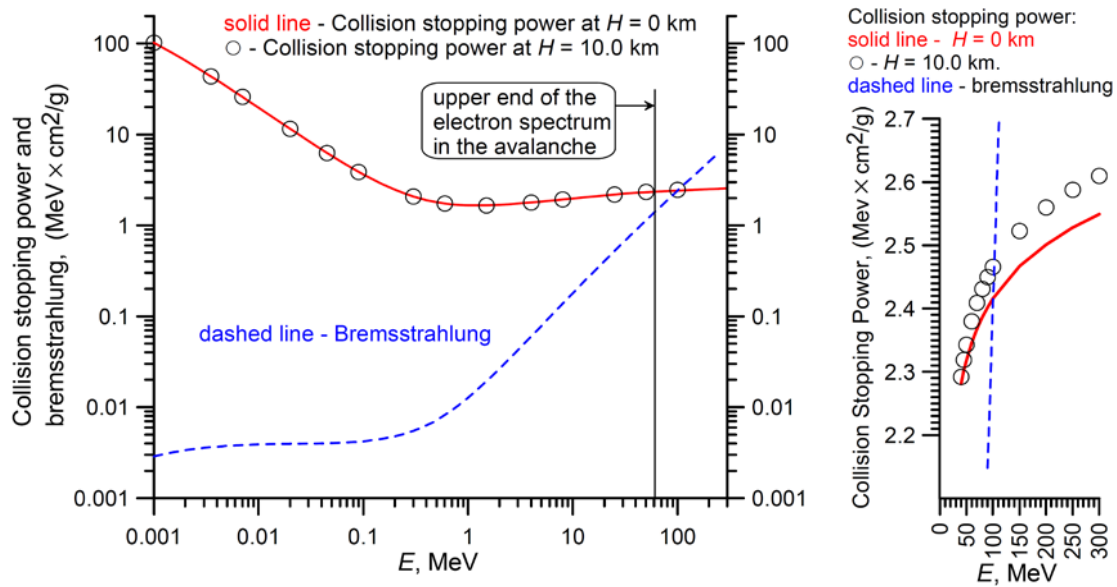

**Figure 8.** Dependence of normalized (on density) collision stopping power and bremsstrahlung on electron energy. The curves for collision stopping power are given for two altitudes $H$ (0 and 10 km). The right part of the figure demonstrates (in scale) the small increase of collision stopping power at elevation relative to the see level.

## 5. Conclusion

It was considered the synthesis of radiocarbon $^{12}$C creation under the condition of thunderstorm flashes. We propose the gross model to evaluate the upper limit of isotope $^{14}$C creation. The synthesis of $^{14}$C is ensured by relativistic electrons in the flash avalanche. The yield of relativistic electrons in the total number of electrons is strongly model dependent. In case $N_{le}/N_{re} \approx 1.3 \times 10^4$ (relation of low energy to relativistic electrons number) [4] the upper limit of $^{14}$C creation on the Earth per year during the thunderstorm is evaluated as $5 \times 10^{-4}$ (g-mole/year). Compare to $^{14}$C isotope creation from the cosmogenic radiation (472 g-mole/year [1]) the production at the thunderstorm gives about $1 \times 10^{-4}$ %. In case of realization of the alternative avalanche model [9] the relative number of relativistic electrons in the avalanche will be in two orders smaller (according to this theory the relation $N_{le}/N_{re} \approx 3 \times 10^6$) and $^{14}$C synthesis on the Earth during the thunderstorm per year will be in two orders smaller too.
It was also shown that dependence of bremsstrahlung yield (the bremsstrahlung is responsible for generation of neutron flux in (γ,n) reaction and generation of $^{14}$C under neutron irradiation) on the air density (with change of the altitude) is very small.

**Acknowledgements**
The author tender thanks to I. N. Borzov for helpful and useful discussion.